\documentstyle[12pt]{article}
\newcommand{\beqn}{\begin{equation}}
\newcommand{\eeqn}{\end{equation}}
\newcommand{\beqnar}{\begin{eqnarray}}
\newcommand{\eeqnar}{\end{eqnarray}}
\newcounter{abc}
\newcommand{\beqna}%

{\renewcommand{\theequation}{\arabic{section}.\arabic{equation}.\alph{abc}}
     \setcounter{abc}{1}
     \begin{equation}}
\newcommand{\eeqna}{\end{equation}
     \renewcommand{\theequation}{\arabic{section}.\arabic{equation}}}
\newcommand{\beqnb}%

{\renewcommand{\theequation}{\arabic{section}.\arabic{equation}.\alph{abc}}
     \setcounter{abc}{2}
     \begin{equation}}
\newcommand{\eeqnb}{\end{equation}
     \renewcommand{\theequation}{\arabic{section}.\arabic{equation}}}
\newcommand{\beqnc}%

{\renewcommand{\theequation}{\arabic{section}.\arabic{equation}.\alph{abc}}
     \setcounter{abc}{3}
     \begin{equation}}
\newcommand{\eeqnc}{\end{equation}
     \renewcommand{\theequation}{\arabic{section}.\arabic{equation}}}
\newcommand{\beqnd}%

{\renewcommand{\theequation}{\arabic{section}.\arabic{equation}.\alph{abc}}
     \setcounter{abc}{4}
     \begin{equation}}
\newcommand{\eeqnd}{\end{equation}
     \renewcommand{\theequation}{\arabic{section}.\arabic{equation}}}
\newcommand{\beqne}%

{\renewcommand{\theequation}{\arabic{section}.\arabic{equation}.\alph{abc}}
     \setcounter{abc}{5}
     \begin{equation}}
\newcommand{\eeqne}{\end{equation}
     \renewcommand{\theequation}{\arabic{section}.\arabic{equation}}}
\newcommand{\beqnarlett}{
     \renewcommand{\theequation}{\arabic{section}.\arabic{equation}.\alph{abc}}
     \begin{eqnarray}}
\newcommand{\eeqnarlett}{\end{eqnarray}
     \renewcommand{\theequation}{\arabic{section}.\arabic{equation}}}
\begin{document}
\titlepage
\setcounter{page}{0}
\vspace{4.5cm}
\title{Off-Shell Bethe Ansatz Equation and N-point Correlators
in the $SU(2) WZNW$ Theory}
\author{H. M. Babujian$^{\ast \dagger}$}
\date{}
\maketitle
\begin{abstract}
We prove that the wave vectors of the off-shell Bethe Ansatz equation for the
inhomogeneous $SU(2)$ lattice vertex model render in the quasiclassical limit
the solutions of the Knizhnik-Zamolodchikov equation.
\\
\\
\\
$^\ast$ Physikalisches Institut der Universit\"at Bonn, Nussallee 12, \newline
D-53115 Bonn, Germany.
\\
Permanent address: Yerevan Physics Institute, \newline
Alikhanian Brothers 2, Yerevan, 375036 Armenia
\\
\\
$^\dagger$
Partially supported by the grant 211-5291 YPI of the German Bundesministerium
f\"ur Forschung und Technologie
\end{abstract}
\newpage
\renewcommand{\theequation}{\arabic{section}.\arabic{equation}}
\section{Introduction}
There exists a rather large class of integrable vertex models in 2d
statistical mechanics, and among them many are gapless. The long range
behavior of these gapless models is described by conformal field theories
(CFT). The finite size resolution of the Bethe Ansatz equations provides
the values of the central charge and the conformal dimensions for the CFT's
corresponding to integrable vertex models [1]. It is well known, that the
Yang-Baxter equation plays the central role in constructing an integrable
vertex model in 2d statistical mechanics. With each simple Lie algebra is
associated a solution of the Yang-Baxter equation and therewith is given an
integrable vertex model. Most results so far obtained are related to
homogeneous vertex models. We will consider in this note instead inhomogeneous
vertex model. Associated is in the case of inhomogeneous models with each
vertex besides the spectral parameter $\lambda$ also a disorder parameter $z$
(one for each side). The vertex weight matrix $R$ depends on $\lambda - z$. The
transfer-matrix of the vertex model so depends now on disorder parameter
$z_i , i = 1, 2 \cdots N$. Transfer-matrices with different values of spectral
parameter $\lambda$ commute which each other $[ 2 , 3 , 4 ]$, which means
that the models are integrable (sect. 2). The purpose of this article is to
investigate the connection between the integrable inhomogeneous vertex
model and conformal field theory. The main ingredient of our approach will be
the wave vector $\Phi ( \lambda_1  \lambda_2 \cdots \lambda_n )$ of the
Algebraic Bethe Ansatz satisfying by construction an equation of the form [5]
\beqn
T ( \lambda ) \Phi (\lambda_1  \lambda_2 \cdots \lambda_n) \; = \;
\Lambda (\lambda , \lambda_1 ,  \lambda_2 \cdots \lambda_n) \; \Phi \;
(\lambda_1 , \cdots \lambda_n) - \sum^{n}_{\alpha = 1}
\frac{F_\alpha \Phi_\alpha}{\lambda - \lambda_\alpha}
\eeqn
Here $T (\lambda)$ denotes the transfer matrix of the vertex model acting on
a $N$ fold tensor product of $SU(2)$ representation spaces.

Where
$\Phi_\alpha = \Phi (\lambda_1 , \cdots \lambda_{\alpha - 1} , \lambda ,
\lambda_{\alpha + 1} ,\cdots \lambda_n) , $
i. e. in $\Phi_\alpha  \; \lambda_\alpha$ is replaced by $\lambda$.
$ F_\alpha (\lambda_1 , \cdots \lambda_n)$ and
$\Lambda (\lambda , \lambda_1 \cdots \lambda_n)$ are $c$ numbers (sect. 2).
The vanishing of the so called "unwanted terms" (the last term on the
r. h. s. of eq. (1.1)) is enforced in the usual procedure of the Bethe Ansatz
by choosing the parameters $\lambda_1 ,  \cdots \lambda_n$ s. t. the
functions $F_\alpha$ vanish. $\Phi$  becomes then eigenvector of the transfer
matrix with eigenvalue $\Lambda (\lambda , \lambda_1 \cdots \lambda_n)$.
We will however not impose these "mass sheel" conditions.
For us the "unwanted" terms are wanted.
We call eq. (1.1) the Off-Shell-Bethe-Ansatz Equation (OSBAE).
Note that all objects in the OSBAE (1.1) depend on the disorder parameters
$z_1 , \cdots z_N$. Our main purpose in this article is to uncover a neat
relationship between the wave vectors satisfying the OSBAE (1.1) and vector
valued solutions of the Knizhnik-Zamolodchikov
(KZ) equation. The latter linear differential equation is of the form [6]
\beqn
\kappa \frac{d \psi}{d z_j} = \sum^N_{i \neq j}
\frac{t^a_j t^a_i}{z_j - z_i} \psi
\eeqn
The variables $z_1 , \cdots z_N$ will be related to the disorder parameters
of the Bethe Ansatz. $ \psi (z_1 , \cdots z_N)$ is a vector in the tensor
product of spaces $V^{(1)} \otimes V^{(2)} \otimes \cdots V^{(N)}$ , where
$V^{(i)} i = 1 , \cdots N$ are representation spaces of the simple algebra $g$.
The $t^a_i (a = 1 , 2 , \cdots \dim g )$ represent the hermitian
generators of the algebra $g$ and act nontrivially on
$V^{(i)}, \kappa = \frac{1}{2 (C_v + K )}$
and
$\delta^{ab} C_V = f^{a c d } f^{b c d}$ ($f^{a b c}$ denoting the structure
constants of the algebra g). $K$ is the central charge of the Kac-Moody
algebra. In this article we consider equation (1.2) only for the group SU(2).
The starting point of our work is the Yang-Baxter equation. We construct
inhomogeneous vertex models with the disorder parameters $\{ z_i \}$ using
SU(2) invariant rational solutions of the Yang-Baxter equation (sect. 2).
The technique of the Algebraic Bethe Ansatz will allow us to find vectors
$\Phi ( \lambda_1 , \cdots \lambda_n )$ satisfying eq. (1.1) (sect. 2). The
solution of the Yang-Baxter equation and OSBAE depend on a parameter $\eta$
(Planck type constant). Sect. 3 is devoted to a discussion of the
quasiclassical limit; $\eta \rightarrow 0$, of the OSBAE which is identified
with a spin wave problem treated by Gaudin [11] some time ago. In sect. 4
we construct from $\Phi ( \lambda_1 , \cdots \lambda_n )$ solution of the KZ
equation. This article is a revised version of ref. [7], which has been
circulated three years ago. The identification of the Gaudin spin problem
with the quasiclassical limit of the general Bethe Ansatz problem is a new
result.
\section{Inhomogeneous vertex model}
\setcounter{equation}{0}
Let us consider a two-dimensional $M \times N$ lattice with $N + 1$ in
general different types of spin variables placed in the following manner
inhomogeneously on the links of the lattice; on all horizontal links are spin
variables $\sigma$ taking values $\pm 1/2$. The variables in the $j - th $
column $j = 1 , 2 , \cdots N$ take values of an SU(2) representation with
spin $s_j$. The interaction takes place only between spins located on
neighboring links and is described by the vertex weight matrix
$R^{j_1 j_2}_{i_1 i_2} (\lambda - z)$, $\lambda$ here is the usual spectral
parameter. $z$ is a local
disorder parameter associated with the particular bond. Cyclic boundary
conditions are imposed. We use the SU(2) invariant solution of the Yang-Baxter
equation [7]
\beqn
_\sigma R^{12} (\lambda - \mu) _S R^{13}
 (\lambda - z)  _S R^{23} (\mu - z) \; = \;
_S R^{23} (\mu - z) _S R^{13} (\lambda - z) _\sigma R^{12} (\lambda - \mu)
\eeqn
where $_\sigma R^{12} (\lambda)$ is the vertex weight matrix of the
$XXX$-model with spin $1/2$ [8],
\beqn
_\sigma R^{12} (\lambda) = I^1 \otimes I^2 + \frac{2 \eta}{\eta - 2 \lambda}
\vec{\sigma}_1 \otimes \vec{\sigma}_2
\eeqn
and
\beqn
_S R^{12} (\lambda - z) = I^1 \otimes I^2 +
\frac{2 \eta}{\eta - 2 (\lambda - z)}
\vec{\sigma}_1 \otimes \vec{S}_2
\eeqn
$\vec{\sigma} = (\sigma^1 , \sigma^2 , \sigma^3 )$ are Pauli matrices,
$\vec{S} = (S^1 , S^2 , S^3 )$ denotes
an operator of arbitrary spin, $I^1 , I^2$ are unit operators in the respective
representation spaces (corresponding to spin $1/2$ and $s$ respectively). The
solution (2.3) has been used in connection with the Kondo problem [9] and
also in the exact solution of the $XXX$-model with arbitrary spin [7]. The
parameter $\eta$ in the $R$-matrices (2.2), (2.3) supplies the quasiclassical
expansion
\beqn
R^{12} (\lambda , \eta ) \mid_{\eta = 0}  = I^1 \otimes I^2
\eeqn
The monodromy operator and transfer-matrix are given in terms of the
R-matrices (2.3) by
\beqnar
J (\lambda , z) = R^{ON} (\lambda - z_N) R^{ON - 1} (\lambda - z_{N - 1})
\cdots R^{0 1} (\lambda - z_1) \nonumber \\
T (\lambda , z) = tr_0 J(\lambda , z )
\eeqnar
In (2.5) the trace is taken in the horizontal two-dimensional space $0$ and
\beqn
R^{0k} (\lambda - z_k ) = I^0 \otimes I^k +
\frac{2 \eta}{\eta - 2 (\lambda - Z_k)}  \;
\vec{\sigma}_0 \otimes \vec{S}_k
\eeqn
operators $\vec{S}_k$ act in the vertical space $V^{(k)}$ and
$(\vec{S}_k)^2 = s_k (s_{k + 1} )$. One infers from eq. (2.1) for the monodromy
operator $J ( \lambda , z )$ the relation
\beqnar
_\sigma R^{12} ( \lambda - \mu ) (J^1 (\lambda , z) \otimes J^2 (\mu , z ))
\; = \:
\nonumber \\
( J^2 (\mu , z ) \otimes J^1 (\lambda , z)) _\sigma R^{12} (\lambda - \mu) .
\eeqnar
Due to the fact, that $T ( \lambda , z ) = t_{r 0} J (\lambda , z )$
we have a family of commuting fransfer matrices as in the homogeneous case
\beqn
\left[ T (\lambda , z) , T (\mu , z) \right] = 0
\eeqn
It is possible to diagonalize the transfer matrix $T ( \lambda , z)$ by the
Algebraic Bethe Ansatz [5], in just the same way as in the homogeneous case
[9,7]. Here we describe this diagonalization. Let the operators
$A ( \lambda , z ) ,
 B ( \lambda , z ) $, $C ( \lambda , z ) , D ( \lambda , z )$
be given by
\beqn
J ( \lambda , z ) =
{A ( \lambda , z ) \; B ( \lambda , z )
 \choose C ( \lambda , z ) \; D ( \lambda , z )}
\eeqn
The matrix $_\sigma R (\lambda )$ in eq. (2.7) can be represented as
\beqnar
_\sigma \dot{R} ( \lambda ) = \begin{array}{c}
\left( \begin{array}{cccc}
		1 & 0 & 0 & 0  \\
		0 & c & b & 0  \\
		0 & b & c & 0  \\
		0 & 0 & 0 & 1   \end{array} \right) ,
\end{array}
\begin{array}{c}
b (\lambda) = \frac{\eta}{\eta - \lambda} \nonumber \\
c (\lambda) = \frac{\lambda}{\lambda - \eta} ,
\end{array}
\eeqnar
Eq. (2.7) gives us the commutators between the elements of $J (\lambda , z).$
We write down the commutators we will use later on
\beqnar
\left[ A (\lambda , z) , A (\mu , z) \right] & = &
\left[ D (\lambda , z) , D (\mu , z) \right] = 0 \nonumber \\
\left[ B (\lambda , z) , B (\mu , z) \right] & = &
\left[ C (\lambda , z) , C (\mu , z) \right] = 0
\eeqnar
\beqnar
B (\lambda , z) , A (\mu , z) = b (\lambda - \mu ) B (\mu , z )
A (\lambda , z ) + c (\lambda - \mu) A (\mu , z) B (\lambda , z)
\eeqnar
\beqnar
B (\mu , z ) D(\lambda , z) = b (\lambda - \mu) B (\lambda , z )
D(\mu , z) + c (\lambda - \mu) D(\lambda , z) B(\mu , z)
\eeqnar
Let us consider the highest weight vector
\beqnar
\mid \Omega > = \mid s_1 , s_1 > \otimes \mid s_2 , s_2 > \otimes
\cdots \otimes \mid s_N , s_N > \nonumber \\
S^3_i \mid s_i , s_i > = s_i \mid s_i , s_i >
\eeqnar
We have the following well known relations for the elements of the
monodromy matrix
\beqnar
A ( \lambda , z ) \mid \Omega > & = &
\prod^N_{i = 1} ( 1 + P_i (\lambda) s_i) \mid \Omega > \\
D ( \lambda , z ) \mid \Omega > & = &
\prod^N_{i = 1} ( 1 - P_i (\lambda) s_i) \mid \Omega > \\
C ( \lambda , z ) \mid \Omega > & = & 0
\eeqnar
In Eq. (2.15) and (2.16)
$P_i (\lambda) = \frac{2 \eta}{\eta - 2 (\lambda - z_i)}$.
The Bethe wave function is
\beqn
\Phi (\lambda_1 , \lambda_2 , \cdots \lambda_n  \; ; \: \{ z \} ) \; = \;
\prod^n_{\alpha = 1} B (\lambda_\alpha , z ) \mid \Omega >
\eeqn
Using Eq. (2.12) - (2.13) and (2.15) - (2.17), we find the action of the
transfer matrix
$T(\lambda , z) = A ( \lambda , z ) + D (\lambda , z )$ on the Bethe vector
$\Phi$
\beqn
T ( \lambda , z ) \Phi = \Lambda (\lambda , \lambda_1 , \lambda_2 , \cdots
\lambda_n ) \Phi - \sum^n_{\alpha = 1}
\frac{F_\alpha (\lambda , z )}{\lambda - \lambda_\alpha} \Phi_\alpha
\eeqn
where
\beqnar
\Lambda (\lambda , \lambda_1 , \lambda_2 , \cdots \lambda_n ) \; = \;
\prod^N_{i = 1} (1 + P_i (\lambda) s_i ) \prod^N_{\alpha = 1}
\frac{1}{c (\lambda_\alpha - \lambda)} + \nonumber \\
+ \prod^N_{i = 1} ( 1 - P_i (\lambda ) s_i ) \prod^n_{\alpha = 1}
\frac{1}{c (\lambda - \lambda_\alpha)}
\eeqnar
\beqnar
F_\alpha (\lambda , z ) = \eta \prod^N_{i = 1} ( 1 + P_i (\lambda) s_i )
\prod^n_{\beta \neq \alpha}
\frac{\lambda_\alpha - \lambda_\beta + \eta}{\lambda_\alpha - \lambda_\beta} -
\nonumber \\
-\eta \prod^N_{i = 1} ( 1 - P_i (\lambda) s_i )
\prod^n_{\beta \neq \alpha}
\frac{\lambda_\alpha - \lambda_\beta - \eta}{\lambda_\alpha - \lambda_\beta}
\eeqnar
\beqn
\Phi_\alpha = \Phi ( \lambda_1 , \lambda_2 , \cdots \lambda_{\alpha - 1} ,
 \lambda , \lambda_{\alpha + 1} , \cdots \lambda_n )
\eeqn
In Eq. (2.22) $\lambda_\alpha$ is replaced by $\lambda$ in $\Phi_\alpha$
or in other words $B (\lambda_\alpha , z)$ is replaced by $B (\lambda , z)$.
The next
step in the traditional Bethe Ansatz procedure would consist in imposing the
vanishing of the so called "unwanted" term (the second term in (2.19)). This
is achieved through an appropriate choice of the parameters
$\lambda_1 , \cdots \lambda_n$ s.t. the funcitons $F_\alpha$ vanishes.
One arrives the true eigenvalue equation
\beqn
T(\lambda , z ) \Phi = \Lambda (\lambda , \lambda_1 , \lambda_2 , \cdots
\lambda_n ) \Phi
\eeqn
The Bethe Ansatz equations $F_\alpha = 0 $ classify the eigenvectors and
eigenvalues of the operator $T (\lambda , z)$. One can say that, when the Bethe
Ansatz equations $F_\alpha = 0$ are satisfied, the Bethe wave
function is on the mass shell. If the condition $F_\alpha = 0$ is not
imposed, then, in general, we have the equation (2.19), and the Bethe wave
function is off the mass shell. In this case we call equation (2.19) the Off-
Shell Bethe Ansatz Equation (OSBAE).
\section{The quasiclassical limit of OSBAE and nonlocal Gaudin Hamiltonians}
\setcounter{equation}{0}
By quasiclassical expansion one commonly understands the expansion of the
vertex weight $R (\lambda , \eta )$ around some point $\eta_0$, such, that
$R (\lambda , \eta_0 ) = I \otimes I [10]$. In this case one can
parameterize $\eta$ so that $\eta_0 = 0$. In this section we calculate the
quasiclassical  limit of the OSBAE (2.19). Let us start from calculation of
the $T (\lambda , z )$ which beside the parameters $\lambda , z_i$ depends also
on $\eta $. In the homogeneous case $(z_i = 0)$ we have for the rational
solution of the Yang-Baxter equation essentially the same structure in the
limits $\eta \rightarrow 0$ and $\lambda \rightarrow \infty$. In the
inhomogeneous case we have another situation, because the dependence on
$z_i$ is additive with $\lambda$.

For the power series expansion $T(\lambda , z )$ around the point
$\eta = 0 $
\beqn
T (\lambda , z ) = \sum^\infty_{k = 0} \eta^k T_k ( \lambda , z)
\eeqn
follows from eq. (2.8) that
\beqn
\sum_{k + m = l} \left[ T_k (\lambda , z ), T_m (\mu , z) \right] = 0 \quad
l, k, m = 0 , 1 , 2 \cdots
\eeqn
which means the existence of the integrable subsystem in the quasiclassical
series (3.1). It is interesting to note that the operators $T_k (\lambda , z )$
in general do not commute wiht $T (\lambda , z )$. In order to find
quasiclassical expansion, we represent the $R$-matrices (2.6) in the
following form:
\beqn
R^{0i} (\lambda - z_i ) =
{I + P_i (\lambda ) S^3_i  \; , \; P_i (\lambda ) S^-_i \choose
P_i (\lambda ) S^+_i \; \; 1 - P_i (\lambda ) S^3_i }
\eeqn
At $\eta < < 1$ we have
\beqn
P_i (\lambda) = - \frac{\eta}{\lambda - z_i} - \frac{1}{2}
\left( \frac{\eta}{\lambda - z_i} \right)^2
\eeqn
Substituting the first term from (3.5) into (3.3), we find the classical
$r$-matrix [10]. From the Eq. (3.3) and (3.4) we obtain the quasiclassical
expansion of the monodromy operatoros $J (\lambda , z)$
(with accuracy $0 (\eta^3))$
\beqn
J (\lambda , z )
 = \prod^N_{i = 1} \left\{ I_0 \otimes I_i \right. + P_i (\lambda )
\left. { S^3_i \quad S^-_i \choose S^+_i \quad -S^3_i} \right\}
\eeqn
Simple calculations give us
\beqnar
A (\lambda , z ) = I - \eta S^3 (\lambda , z) + \eta^2 \sum_{i < i}
\frac{S^3_i S^3_j + S^-_i S^+_j}{(\lambda - z_i) (\lambda - z_j)} +
\nonumber \\
+ \frac{\eta^2}{2} \frac{d}{d \lambda} S^3 (\lambda , z )
+ 0 (\eta^3 )
\eeqnar
\beqnar
D (\lambda , z ) = I - \eta S^3 (\lambda , z) + \eta^2 \sum_{i < i}
\frac{S^3_i S^3_j + S^+_i S^-_j}{(\lambda - z_i) (\lambda - z_j)}  -
\nonumber \\
- \frac{\eta^2}{2} \frac{d}{d \lambda} S^3 (\lambda , z )
+ 0 (\eta^3 )
\eeqnar
\beqnar
B (\lambda , z ) =  - \eta S^- (\lambda , z) + \eta^2 \sum_{i < i}
\frac{S^3_i S^-_j - S^-_i S^3_j}{(\lambda - z_i) (\lambda - z_j)} -
\nonumber \\
- \frac{\eta^2}{2} \frac{d}{d \lambda} S^- (\lambda , z )
+ 0 (\eta^3 )
\eeqnar
\beqnar
C (\lambda , z ) = - \eta S^+ (\lambda , z) + \eta^2 \sum_{i < i}
\frac{S^+_i S^3_j - S^3_i S^+_j}{(\lambda - z_i) (\lambda - z_j)} -
\nonumber \\
- \frac{\eta^2}{2} \frac{d}{d \lambda} S^+ (\lambda , z )
+ 0 (\eta^3 )
\eeqnar
In Eqs. (3.6) - (3.9) $I = \prod^N_{i = 1} \otimes I_i$, and also we use the
following notation
\beqn
S^a (\lambda , z ) = \sum^N_{i = 1} \frac{S^a_i}{\lambda - z_i}
\eeqn
where
$S^a = (S^1 , S^2 , S^3 )$ and
$S^\pm (\lambda , z ) = S^1 (\lambda , z ) \pm i S^2 (\lambda , z)$.
For the transfer matrix $T (\lambda , z ) = A (\lambda , z ) + D (\lambda z )$
we have
\beqn
T (\lambda , z) = 2 I + 2 \eta^2 \sum^N_{j = 1} \frac{H_j}{\lambda - z_j}
\eeqn
\beqn
H_i = \sum^N_{i \neq j} \frac{S^a_j S^a_i}{z_j - z_i}
\eeqn
One can see from (3.1) and (3.11) that
$T_0 (\lambda , z) = 2 I , T_1 (\lambda , z ) = 0$ and second term in (3.11) is
equal to $T_2 (\lambda , z)$. It is obvious that the operators $H_j$ commute
as consequence eq. (3.2). Then we can calculate with (3.6) - (3.9) the
quasiclassical limits of the objects (2.20) - (2.22):
\beqn
\Phi (\lambda_1 , \lambda_2 , \cdots \lambda_n ) =
(-\eta)^n \prod^n_{\alpha = 1} S^- ( \lambda_\alpha , z ) \mid \Omega > +
0 (\eta^{n + 1})
\eeqn
\beqnar
\Lambda (\lambda , \lambda_1 , \lambda_2 , \cdots \lambda_n ) = 2 + 2 \eta^2
\left\{ \sum_{i \alpha} \right.
\frac{s_i}{(\lambda - z_i ) (\lambda_\alpha - \lambda)} +
\nonumber \\
+ \sum_{i \neq j} \frac{s_i s_j}{(\lambda - z_i ) (\lambda - z_j)} +
 \sum_{\alpha \neq \beta}
\left. \frac{1}{(\lambda_\alpha - \lambda ) (\lambda_\beta - \lambda)} \right\}
+ 0 (\eta^3)
\eeqnar
\beqn
F_\alpha = 2 \eta^2 \left\{ \sum^n_{\alpha \neq \beta}
 \frac{1}{\lambda_\alpha - \lambda_\beta} \right.
- \left.  \sum^N_{i = 1} \frac{S_i}{\lambda_\alpha - z_i} \right\} + 0 (\eta^3
)
\eeqn
\beqnar
\Phi_\alpha = ( - \eta )^n S^- (\lambda_1 , z ) \cdots S^-
( \lambda_{\alpha - 1} , z ) S^- ( \lambda , z) S^- ( \lambda_{\alpha + 1} , z)
 \cdots
\nonumber \\
S^- ( \lambda_n , z ) \times \mid \Omega > + 0 (\eta^{n + 1})
\eeqnar
Substituting now Eqs. (3.13) - (3.16) and (3.11) into (2.19) and combining
the terms proportional to $\eta^{n + 2}$, we obtain the first nontrivial
consequence of OSBAE (2.19) in the quasiclassical limit
\beqn
\sum^N_{j = 1} \frac{H_j}{\lambda - z_j} \varphi = h \varphi -
\sum^n_{\alpha = 1} \frac{f_\alpha \varphi_\alpha }{\lambda - \lambda_\alpha}
\eeqn
Here the vectors $\varphi$ and $\varphi_\alpha$ from (3.14) and (3.16) are
propotional to $\eta^2$ in (3.14), (3.15), respectively. Taking the residue
in the pole $\lambda = z_j$ of the Eq. (3.17) we have
\beqn
H_j \varphi = h_j \varphi \; - \; \sum^n_{\alpha = 1}
\frac{f_\alpha S^-_j}{z_j - \lambda_\alpha} \varphi^\prime_\alpha
\eeqn
where
\beqn
h_j = \sum^N_{j \neq j}  \frac{s_i s_j}{z_j - z_i} \; + \;
\sum^n_{\alpha = 1} \frac{s_i}{\lambda_\alpha - z_j}
\eeqn
\beqn
f_\alpha =
 \sum^n_{\beta \neq \alpha} \frac{1}{\lambda_\alpha - \lambda_\beta} \; - \;
\sum^N_{i = 1} \frac{s_i}{\lambda_\alpha - z_i}
\eeqn
\beqn
\varphi (\lambda , z) = \prod^n_{\alpha = 1} S^- (\lambda_\alpha , z )
\mid \Omega >
\eeqn
In (3.18) we define the vector
$\varphi^\prime_\alpha ; \varphi
 = S^- (\lambda_\alpha , z ) \varphi^\prime_\alpha$
i. e. in the vector $\varphi^\prime_\alpha$ the operator
 $S^- (\lambda_\alpha , z ) $
is omited. Eqs. (3.18) and (3.19) - (3.21) reproduce Gaudin's results [11],
which he found considering the spectral problem for the set of operators $H_j$.
{}From Eqs. (3.18) - (3.21) it also follows, that the Gaudin method is in fact
a quasiclassical version of the Algebraic Bethe Ansatz. If in (3.18) we impose
the condition $f_\alpha = 0$, then we obtain $\varphi$ as eigenvector of
the operators $H_j$ with eigenvalues $h_j$. Parameters
$\lambda_1 , \cdots \lambda_n$ have to be found from the quasiclassical Bethe
Ansatz equations $f_\alpha = 0$.
\section{The integral representation for the N-Point correlators in WZNW
theory}
\setcounter{equation}{0}
Let us introduce the function
$\chi ( \lambda , z ) = \chi (\lambda_1, \cdots \lambda_n , z_1 \cdots z_N)$
obeying the following differential relations [12]
\beqn
\kappa \frac{d \chi}{d z_j} = h_j \chi
\eeqn
\beqn
\kappa \frac{d \chi}{d \lambda_\alpha} = f_\alpha \chi
\eeqn
Where $\kappa = \frac{1}{2 (k + 2)}$ (in the case $SU(2) C_v = 2$).
taking into account Eqs (3.19), (3.20) it is easy to verify that the zero
curvature conditions are fulfilled
$\frac{d h_j}{d \lambda_\alpha} = \frac{d f_\alpha}{d z_j}$.

The solution of the Eqs. (4.1) - (4.2) is
\beqnar
\chi (\lambda , z) = \prod^N_{i < j} (z_i - z_j )^{\frac{S_i S_j}{\kappa}}
\prod^n_{\alpha < \beta} (\lambda_\alpha - \lambda_\beta )^{\frac{1}{\kappa}}
\nonumber \\
\prod_{k \gamma} (z_k - \lambda_\gamma )^{- \frac{S_k}{\kappa}}
\eeqnar
We define vector-function $\Psi (z_1 \cdots z_N )$ through multiple contour
integrals as follows [12]
\beqn
\Psi (z_1 , \cdots z_N ) = \oint \cdots \oint \chi (\lambda , z )
\varphi (\lambda , z ) d \lambda_1 , \cdots d \lambda_n
\eeqn
The integrations are to be taken here over canonical cycles of the space
$X = C^n - U (\lambda_\alpha = z_i )$ with coefficients from $S^\ast_\lambda $
dual to the local system $S_\lambda $, that is defined by the monodromy group
of the function $\chi ( \lambda , z )$. It is now rather straightforward to
show that vector-function $\Psi ( z_1  \cdots z_N )$ defined above is the
solution of KZ Eq. (1.2). Substituting (4.5) into KZ equation (1.2), using
OSBAE (3.18) and the defining relations for $\chi$ (4.1) we find
\beqn
\oint \cdots \oint \left[ \chi \frac{d \varphi}{d z_j} - \frac{1}{\kappa}
\sum^n_{\alpha = 1}
\frac{S^-_j f_\alpha \chi \varphi^\prime_\alpha }{z_j - \lambda_\alpha} \right]
d \lambda_1 , \cdots d \lambda_n = 0
\eeqn
Taking into account (4.2) and the identity which follows directly from (3.21)
\beqn
\frac{d \varphi}{d z_j} = - \sum^n_{\alpha = 1} \frac{d}{d \lambda_\alpha}
\frac{S^-_j \varphi^\prime_\alpha}{z_j - \lambda_\alpha}
\eeqn
one easily verifies that (4.6) boils down to the relation
\[
\sum^n_{\alpha = 1} \oint \cdots \oint \frac{d}{d \lambda_\alpha}
\left[ \frac{S^-_j \varphi^\prime_\alpha \chi}{z_j - \lambda_\alpha} \right]
d\lambda_1 \cdots d \lambda_n = 0
\]
It is evident that this equation is satisfied, because the contours are closed.
Now we want to show that $\Psi (z_1 \cdots z_N)$ is a singlet with respect to
the global $SU(2)$ [6]
\beqn
S^3 \Psi = \sum^N_{i = 1} S^3_i \Psi = 0
\eeqn
\beqn
S^\pm \Psi = \sum^N_{i = 1} S^\pm_i \Psi = 0
\eeqn
Eq. (4.7) can easily be verify if we take into account the relation
\beqn
\left[ S^3 , S^- (\lambda , z ) \right] = - S^- ( \lambda , z )
\eeqn
We have
\beqn
S^3 \Psi = \oint \cdots \oint \left[ \sum^N_{i = 1} s_i - n \right]
\chi \varphi d \lambda_1 , \cdots d \lambda_n = 0
\eeqn
since we now impose the condition
\[
n =  \sum^N_{i = 1} s_i
\]
In order to verify Eq. (4.8) we present the Bethe wave function $\varphi $
in the correlation functions (4.5) in more explicit form. It is indeed given
by a sum of integrals of the Aomoto-Gelfand type [13, 14]
\beqn
\chi_{k_1 , k_2 \cdots k_n} (z_1 , \cdots z_N ) = \nonumber
\\
\oint \cdots \oint
\frac{\chi (\lambda , z ) d \lambda_1 \cdots d\lambda_n}{(z_{k_1}
 - \lambda_1) (z_{k_2} - \lambda_2 ) \cdots (z_{k_n} - \lambda_n)}
\eeqn
Where $k_\alpha = 1 , 2 \cdots N$. Aomoto had studied such integrals in
connection with general hypergeometric functions. It follows from his work that
we can take such cycles (contours) where the integral will be completely
symmetric under the permutation of any $\lambda_\alpha$. In this case it will
not depend on $k_1 ,\cdots k_n$ in (4.13) i. e. on repetitions of
a given $z$ in the denominator of the integrand. So, we denote these integrals
$\chi_{q_1 \cdots q_n ( z_1 , \cdots z_N )}$; where the $q_i$ are
repetition numbers of given $z$ in the denominator of (4.13). We can represent
$\Psi$ in the form
\beqnar
\Psi (z_1\cdots z_N ) = \nonumber \\
\oint \cdots \oint \chi (\lambda , z ) \sum^N_{k_1 = 1}
\frac{S^-_{k_1}}{(z_{k_1} - \lambda_1)} \cdots \sum^N_{k_n = 1}
\frac{S^-_{k_1}}{z_{k_n} - \lambda_n} \mid \Omega > d \lambda_1 \cdots
d \lambda_n =
\nonumber \\
= n ! \sum_{q_1 + q_2 \cdots q_N = n} \cdot \chi_{q_1 , \cdots q_N}
\frac{(S^-_j)^{q_1}}{q_1 !} \cdots \frac{(S^-_N)^{q_N}}{(q_N) !}
\mid \Omega >
\eeqnar
We have then
\beqnar
S^+\varphi =
\nonumber \\
\sum_{q_1 + q_2 \cdots q_N = n} \sum^N_{j = 1} \chi_{q_1 \cdots q_N}
\frac{(S^-_1)^{q_1}}{(q_1) !} \cdots \frac{S^+_j (S^-_j)^{q_1}}{(q_j) !}
\cdots \frac{(S^-_N)^{q_N}}{(q_N) !} \mid \Omega >
\eeqnar
Taking into account the commutation relation
\[
\left[ S^+ , (S^-)^k  \right] = k (S^-)^{k - 1} ( 2 S^3 + 1 - k)
\]
we obtain
\beqnar
\sum_{q_1 + q_2 + \cdots q_N = n - 1} \sum^N_{j = 1}
\chi_{q_1 \cdots q_j + 1 , \cdots q_N} (2 s_j - q_j) \times
\nonumber \\
\times \frac{(S^-_1 )^{q_1}}{(q_1) !} \cdots \frac{(S^-_N)^{q_N}}{(q_N) !}
\mid \Omega > = 0
\eeqnar
or equivalently
\beqn
\sum^N_{j = 1} \chi_{q_1 \cdots q_j + 1 , \cdots q_N} (2 s_j - q_j ) = 0
\eeqn
This coincides with one of Aomoto's identities for general hypergeometric
functions [13]. The formula (4.14) establishes the relationship between our
approach and of the authors [15 - 18]. In order to calculate specific
correlation functions, as in [6, 19], i. e. the correlation function of
primary fields $\Phi^{S_i}_{m_i} (z_i ) , m_i = -s_i \cdots s_i$
\beqn
\langle \Phi^{s_1}_{m_1} ( z_1) \Phi^{s_2}_{m_2} (z_r) \cdots
\Phi^{s_N}_{m_N}  (z_N)
\eeqn
it is nessesary to multiply $\Psi (z_1 \cdots z_N)$ from the left hand side by
the vector
\[
< m_1 , s_1 \mid \otimes < m_2 , \mid s_2 \mid \otimes \cdots \otimes
< m_N , s_N \mid
\]
where $< m_i , s_i \mid $ is defined as;
\[
< m_i , s_i \mid S^3_i = m_i < m_i , s_i \mid
\]
where we have the relation
\beqnar
\langle \Phi^{S_1}_{m_1} (z_1) \Phi^{S_2}_{m_2} (z_2) \cdots \Phi^{S_N}_{m_M}
(z_N) \rangle =
\nonumber \\
= < m_1 , s_1 \mid \otimes < m_2 , s_2 \mid \otimes \cdots \otimes \langle
m_N , S_N \mid \Psi (z_1 , z_2 \cdots z_N )
\eeqnar
For full conformity with the WZNW theory in equations (4.1) - (4.2) we take
$\kappa = \frac{1}{2 (k + 2)}$. However, it is clear, that our construction
allows us to work with arbitrary $\kappa$. So the inhomogeneous vertex model
with transfer matrix $T (\lambda , z )$ and OSBAE generate the correlators
of the WZNW theory.
\section*{Conclusion and speculation}
Many connections have been established in the last years relating integrable
spin models and 2d CFT. The main result of this paper consists in adding
another link in this direction: the solutions of the rational SU(2) KZ equation
are identified modulo a scalar integrating factor with a Bethe wave vector of
the Algebraic Bethe Ansatz for an inhomogeneous vertex model in quasiclassical
limit. It has to be stressed that the connection exists between the KZ
equation and the Off-Shell Bethe Ansatz Equation. It should be noted that the
quasiclassical expansion can be reinterpreted as high temperature expansion of
lattice vertex model, because the leading term of this expanison correspond
to maximal entrope of the lattice vertex model.

We find it rather likely in considering the structure of OSBAE that general
Bethe wave vector (beyond the quasiclassical limit) will give the solutions of
quantum KZ equations of rational type. Related results for the trigonometric
quantum KZ equations are due to Matsuo [20] and Reshetikhin [21].
\section*{Acknowledgements}
I am grateful for discussions to A. Sedrakyan and R. Pogosian. I am specially
grateful to R. Flume for discussions and comments. I thank the members of the
Physikalisches Institut, where this work has been finished, for hospitality
extended to me.
\section*{References}
\begin{enumerate}
\item H. J. de Vega, Int. J. Mod. Phys. {\bf A4} (1989) 2371
\item R. J. Baxter, {\it "Exactly solved models in statistical mechanics"},
(Academic Press, London, 1982). \\
R. J. Baxter, Studies Appl. Math. {\bf L51}
(1971)
\item Kulisch and N. Yu. Reshitikhin, J. Phys. {\bf A16} (1983) L591
\item H. J. de Vega, Nucl. Phys. {\bf B240} [FS12] (1984) 495
\item L. D. Faddeev, E. K. Sklyanin and L. A. Takhtajan,
Ther. Mat. Fis. {\bf 40}
(1979) 194
\item V. G. Knizhnik and A. B. Zamolodchikov, Nucl. Phys. {\bf B247} (1984) 83
\item H. M. Babujian, Nucl. Phys. {\bf B214} [FS7] (1983) 317
\item R. J. Baxter, Ann. of Phys. {\bf 70} (1972) 323
\item V. A. Fateev and P. B. Wiegmann, Phys. Rev. Lett {\bf 46} (1981) 1955
\item A. A. Belavin and V. G. Drinfeld, Funct. Anal. Apll. {\bf 16} (1982) 159
\item M. Gaudin, J. Phys. (Paris) {\bf 37} (1976) 1087
\item H. M. Babujian in Proced. XXIV Inter. Symp. Ahrenshoop, Zeuten 1990,
Preprint YERPHI-1261 (47)-90
\item K. Aomoto, J. Math. soc. Japan {\bf 39} (1987) 191
\item V. A. Vasilev, I. M. Gelfand and A. V. Zelevinskii, Funct. Anal. Priloz.
{\bf 21} (1987) 19
\item P. Christe and R. Flume, Nucl. Phys. {\bf B282} (1987) 466. \newline
P. Christe, PhD. Thesis, Bonn University (1986)
\item E. Date, M. Jimbo, a. Matsuo and T. Miwa, in Yang-Baxter Equation,
Conformal Invariance and Integrability in Statistical Mechnics and Field
Theory, World Scientific (1989)
\item V. V. Schechtman and A. N. Varchenko, Lett. Mat. Phys. {\bf 20} (1990)
279, Inven. Math. {\bf 106} (1991) 139
\item A. Matsuo, Comm. Math Phys. {\bf 134} (1990) 65
\item A. B. Zamalodchikov, V. A. Fateev, Sov. J. Nucl. Phys.
{\bf V43} (1986) 1031
\item A. Matsuo, Comm. Math. Phys. {\bf 151} (1993) 263
\item N. Reshetikhin, Lett. in Math. Phys. {\bf 26} (1991) 153
\end{enumerate}
\end{document}